\documentclass[longauth]{aa}
\usepackage{epsfig}
\usepackage{natbib}
\bibpunct{(}{)}{;}{a}{}{,}
\usepackage{graphicx}

\def\ms{\hbox{\,ms$^{-1}$}}         
\def\m2s2{\hbox{\,m$^{2}$\,s$^{-2}$}} 
\def\kms{\hbox{\,kms$^{-1}$}}       
\def\vsini{\hbox{$v$\,sin\,$i$}~}      
\def\Msun{\hbox{$M_{\odot}$}}             
\def\Rsun{\hbox{$R_{\odot}$}}
\def\Mjup{\hbox{$M_{\rm Jup} $}}
\def\Rjup{\hbox{$R_{\rm Jup} ~$}}

\def\exo1{\mbox{CoRoT-Exo-1}}

\newcommand{\teff}{T$_{\rm eff}$}
\newcommand{\logg}{log \it g\rm~}

\newcommand{\gmet}{[M/H]}
\begin{document}

\title{Transiting exoplanets from the CoRoT space mission
\subtitle{I - \exo1b: a low-density short-period planet around a G0V star}
\thanks{The CoRoT space mission, launched on Dec. 27th, 2006, was developed and is operated by the CNES, with participation of the Science Program of ESA, ESTEC/RSSD, Austria, Belgium, Brazil, Germany and Spain. Based in part on observations with the SOPHIE spectrograph at Obs. de Haute Provence, France} }

\author{Barge, P. \inst{1}
\and Baglin, A. \inst{2}
\and Auvergne, M. \inst{2}
\and Rauer, H. \inst{3,20}
\and L\'eger, A. \inst{4}
\and Schneider, J. \inst{5}
\and Pont, F. \inst{6}
\and Aigrain, S. \inst{7}
\and Almenara, J-M. \inst{9}
\and Alonso, R. \inst{1}
\and Barbieri, M. \inst{1}
\and Bord\'e, P. \inst{4}
\and Bouchy, F. \inst{8}
\and Deeg, H.J. \inst{9}
\and De la Reza,  \inst{10}
\and Deleuil, M. \inst{1}
\and Dvorak, R. \inst{11}
\and Erikson, A. \inst{3}
\and Fridlund, M. \inst{12}
\and Gillon, M. \inst{6}
\and Gondoin, P. \inst{12}
\and Guillot, T. \inst{13}
\and Hatzes, A. \inst{14}
\and Hebrard, G. \inst{8}
\and Jorda, L. \inst{1}
\and Kabath, P. \inst{3}
\and Lammer, H. \inst{16}
\and Llebaria, A. \inst{1}
\and Loeillet, B. \inst{1}
\and Magain, P. \inst{18}
\and Mazeh, T. \inst{19}
\and Moutou, C. \inst{1}
\and Ollivier, M. \inst{4}
\and P\"atzold, M. \inst{15}
\and Queloz, D. \inst{6}
\and Rouan, D. \inst{2}
\and Shporer, A.   \inst{19}
\and Wuchterl, G. \inst{14}    }

\offprints{\email{pierre.barge@oamp.fr}}

\institute{Laboratoire dÕAstrophysique de Marseille, UMR6110, CNRS/Universit\'e de Provence - B.P8, 13376 Marseille cedex 12, France
\and
LESIA, Observatoire de Paris-Meudon, 5 place Jules Janssen, 92195 Meudon, France
\and
Institute of Planetary Research, DLR, Rutherfordstr. 2, D-12489 Berlin, Germany
\and
Institut dÕAstrophysique Spatiale, Universit\'e Paris XI, F-91405 Orsay, France
\and
LUTH, Observatoire de Paris-Meudon, 5 place Jules Janssen, 92195 Meudon, France
\and
Observatoire de Gen\`eve, Universit\'e de Gen\`eve, 51 chemin des Maillettes, 1290 Sauverny, Switzerland
\and
School of Physics, University of Exeter, Stocker Road, Exeter EX4 4QL, United Kingdom
\and
Institut dÕAstrophysique de Paris, Universit\'e Pierre \& Marie Curie, 98bis Bd Arago, 75014 Paris, France
\and
Instituto de Astrof\'isica de Canarias, E-38205 La Laguna, Tenerife, Spain
\and
Observat\'orio Nacional, Rio de Janeiro, RJ, Brazil
\and
University of Vienna, Institute of Astronomy, T\"urkenschanzstr. 17, A-1180 Vienna, Austria
\and
Research and Scientific Support Department, ESTEC/ESA, 2200 Noordwijk, The Netherlands
\and
Observatoire de la C\^ote d\'Azur, Laboratoire Cassiop\'ee, BP 4229, 06304 Nice Cedex 4, France
\and
Th{\"u}ringer Landessternwarte, Sternwarte 5, Tautenburg 5, D-07778 Tautenburg, Germany
\and
Rheinisches Institut f\"ur Umweltforschung an der Universit\"at  zu K\"oln, Aachener Strasse 209, 50931, Germany
\and
Space Research Institute, Austrian Academy of Science, Schmiedlstr. 6, A-8042 Graz, Austria
\and
University of Vienna, Institute of Astronomy, T\"urkenschanzstr. 17, A-1180 Vienna, Austria
\and
University of Li\`ege, All\'ee du 6 ao\^ut 17, Sart Tilman, Li\`ege 1, Belgium
\and
School of Physics and Astronomy,  Tel Aviv University, Tel Aviv 69978,  Israel
\and
Center for Astronomy and Astrophysics, TU Berlin, Hardenbergstr. 36, 10623 Berlin
}
\date{Received / Accepted }
\abstract {The pioneer space mission for photometric planet searches, CoRoT, steadily monitors about 12,000 stars in each of its fields of view; it is able to detect transit candidates early in the processing of the data and before the end of a run.}{We report the detection of the first planet discovered by CoRoT and characterizing it with the help of follow-up observations.}{Raw data were filtered from outliers and residuals at the orbital period of the satellite. The orbital parameters and the radius of the planet were estimated by best fitting the phase folded light curve with 34 successive transits. Doppler measurements with the SOPHIE spectrograph permitted us to secure the detection and to estimate the planet mass. \thanks{Table \ref{tab:rad} is  available in electronic form at the CDS via anonymous ftp to cdsarc.u-strasbg.fr (130.79.128.5) or the web site.} }
{The accuracy of the data is very high with a dispersion in the 2.17 min binned phase-folded light curve that does not exceed $\sim3.\times 10^{-4}$ in flux unit. The planet orbits a mildly metal-poor G0V star of magnitude V=13.6 in 1.5 days. The estimated mass and radius of the star are 0.95$\pm0.15$\Msun~and 1.11$\pm0.05$\Rsun . We find the planet has a radius of 1.49$\pm0.08$\Rjup, a mass of 1.03 $\pm0.12$\Mjup~, and a particularly low mean density of 0.38 $\pm0.05${\rm g cm}$^{-3}$. }
{}

\titlerunning{CoRoT-Exo-1b}

\authorrunning{Barge et al.}
\maketitle
\section{Introduction}
%
CoRoT is the first space survey dedicated to the photometric search for extrasolar planets. Led by the CNES space agency, CoRoT was born from a joint effort of France, European countries, ESA,  and Brazil \citep{bag06}. The detailed description of the instrument and the mission can be found in a pre-launch book \citep{book06} and in a post-launch paper (Auvergne et al. {\sl in preparation}). The instrument satisfies the initial scientific requirements of the mission with a noise level that matches the photon noise over most of the magnitude range accessible to CoRoT (12-16). The measure of the stellar flux every 512 s results from the piling up on board of 16 individual exposures of 32 s obtained by aperture photometry. The photometric masks on the CCDs are uploaded at the beginning of a run, after being selected from a pre-defined list of patterns. In addition to the target, a mask may contain contaminating stars that are potential sources of noise and false alarms. Reference windows with a square shape were also selected in the non-exposed parts of the CCDs ($5\times 5$ pixels) the offset corrections or in the darkest regions of the CCDs ($10\times10$ pixels) for sky-background corrections. In polar Earth orbit, CoRoT is free of the main limitations of ground-based transit searches. The photometric accuracy is better than 1mmag and the monitoring of the targets is nearly continuous over several months. However, the total duration of a run cannot exceed 6 months to avoid blinding by the Sun and occultation by the Earth. Another inconvenience comes from charged particle impacts at the crossing of the South-Atlantic Anomaly (SAA) as they can produce bright pixels on the CCDs and outliers in the data. This letter reports the detection and characterization of the first planet discovered by CoRoT in its initial run of observation. With a large radius and a low mean density, the planet seems to depart from the distribution of the known transiting planets. The results are based on a preliminary analysis of the raw data that we briefly describe and on ground-based follow-up with the SOPHIE spectrograph. The full analysis of the light curve will be presented elsewhere using the fully reduced data issued from the standard pipeline of corrections.

\section{CoRoT observations}
CoRoT's observation sequence started with a pointing direction close to the anticenter of the Galaxy. This initial run lasted from February 6th to April 2nd 2007 and supplied $\sim$12,000 high accuracy light curves for exoplanet transit search. These light curves are nearly continuous over 55 days with only a small number of gaps that mainly result from the crossing of the SAA (the duty cycle is 92\%). The nominal sampling is 512 s but can be 32 s for $\sim$10\% of the targets (500 out of 5700 in each of the two CCDs). A peculiarity of the CoRoT exoplanet program is the ``alarm mode'', an operational loop between the science team and ``Control \& Command Center'' on ground and the instrument in space. Its goal is to optimize the science return of the mission by identifying transit candidates early in the process, before the end of run and before the data are fully reduced. For these transit candidates a decision can be made to change the rate of the observations from one exposure in 512 s to one exposure in 32 s. The interest for this oversampling is two-fold: (i) to get a better coverage of the transit profile; (ii) to reduce the level of noise by removing outliers from the corrupted exposures. Last but not least, this ``on the stream'' detection makes it possible to start follow-up operations as soon as possible. Thanks to the alarm mode, some transit candidates were identified soon in the raw data and the sampling rate changed correspondingly. For example, the best candidate, \exo1b, found around a G0V star of ${V}$=13.6 (\exo1) with a period of 1.5 days and a depth of 2\%, was submitted to a series of follow-up observations that started at mid run.

\section{Data and preliminary treatments}
The light curve of \exo1 consists of $\sim$68,824 data points for a total duration of 54.72 days. Sampling is 512 s during the first 30 days and 32 s for the rest of the time. Out of 36 detected transits, 20 are found in the part of the light curve with nominal sampling and 16 are found in the oversampled part. Data used for ``alarm'' detection are raw data requiring further processing to get correct estimates of the transit parameters, particularly, the transit depth. We used a number of treatments that focus:  (i) on the background and offset corrections; (ii) on the filtering of the outliers and low frequency residuals produced at SAA crossings. The developed procedures are forerunners of the ones implemented in the reduction pipeline.\\
$-$ Reference windows for the background and the offset corrections can be affected by charged particle impacts at SAA crossing. To detect and eliminate, in these windows, the resulting outliers, we subtracted a moving-median filtered version of the light-curve and discarded the points at distances greater than a few times the dispersion of the residuals. For the background, a constant level was estimated using the reference sky window closest to the target. The background light curve, once filtered from the outliers, is smoothed out with a 190 points width median filter and interpolated into the data points (as the time sampling changed during the observations). By this method the average background level is 11.9 $ {\rm e^{-}px^{-1}s^{-1}}$ (5.58 ADU$ {\rm~px^{-1}s^{-1}}$).  For the offset, we proceeded in the same way as for the background. The camera offset is obtained as the median of 23 reference windows. The offset curve is low-pass filtered with the same median filter as above, but with a width of 500 points. Thus, the median value of the offset is 55.47${\rm e^{-}px^{-1}s^{-1}}$ (25.9 ADU${\rm~px^{-1}s^{-1}}$). To get the total correction, we made the sum of the background and the offset (estimated within the photometric mask) and subtracted the result to the original data. After multiplying by the detector gain, we found the median flux from the star is 221,000${\rm e^{-}}$ per 32 s exposure (640.45${\rm e^{-}px^{-1}s^{-1}}$) with a signal-to-noise ratio of 476.19 .\\
$-$ The \exo1 light curve also contains outliers that are identified and located thanks to a median filtering, in the same way as performed for the background and offset light curves. With such a procedure, 8\% of the data points were identified as outliers and eliminated from the subsequent analysis. The filtered light curve keeps the signature of the satellite orbital period (at 102 min), likely due to residuals of the background corrections. We modeled this effect by estimating, for each orbit of the satellite, the modulation obtained from the 50 closest orbits; the light curve is successively: (i) normalized with the median value at each orbit, (ii) phase-folded at the period of the satellite, (iii) high frequency filtered (using a $128\times128$ points Savitzky-Golay weighted moving average filter, with 4th degree weighting). From the preparatory ExoDat database only two faint contaminants (${V}> 17$) are present within the 61 pixels of \exo1's photometric mask; their contribution to the total flux was neglected.

\section{Folded light curve and best fitting}
\label{sec:fold}
Possibly due to bright pixels inside the photometric mask, the light curve has low frequency $(<1cycle/day)$ residuals that could mar the transit shape and affect the estimate of the star and planet parameters. To minimize such effect we proceeded as follows: (i) around each individual transit (just before and after), a parabola is chosen to fit the data points close to the transit phase, (ii) the fit is high frequency filtered and subtracted to the ensemble of points ranging from the transit orbital phase $-0.1$ to the phase $0.1$, (iii) the 36 transits are re-sorted in orbital phase, using the ephemeris reported in Table \ref{tab:par}. Then, the points are binned in 0.001 phase-wide bins (2.17 min in time unit), and the error bars are estimated as the dispersion of the points inside the bin divided by the square root of the number of points per bin. The median number of points per bin is 316 and the average 1$\sigma$ error bar estimated this way is 0.0003 ($\sim3$ times larger than expected from the photon noise limit). The noise level should be reduced once the pipeline includes jitter corrections.
\begin{figure}
\begin{center}
\epsfig{file=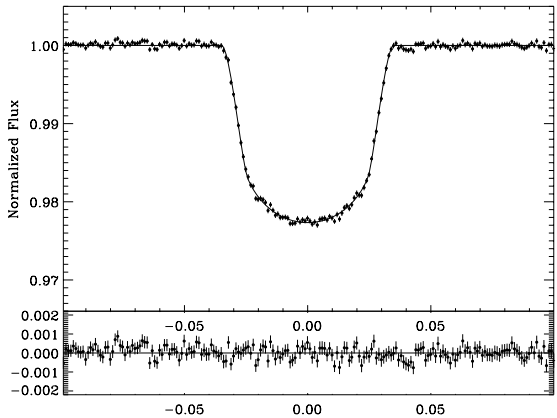,width=8.cm,angle=0}
\caption{Normalized and phase folded light curve of the 34 best transits of \exo1b (top); the residuals from the best-fit model (bottom). The bin size corresponds to $2.17 min$, and the $1\sigma$ error bars have been estimated from the dispersion of the data points inside each bin.}
\label{fig:fig1}
\end{center}
\end{figure}
The detrended light curve was phase-folded, avoiding 2 transits strongly affected by a bright pixel (see Figure \ref{fig:fig1}), and fitted to a model following the formalism of  Gim\'enez (2006). To find the solution that best matches the data, we minimized the $\chi^2$  using the algorithm AMOEBA  \citep{pre92}. The fitted parameters are the transit center, the orbital phase at transit start, $R_{P}/R_{\star}$, the orbital inclination $i$ and the two non linear limb-darkening coefficients $u_{+}$ and $u_{-}$ \citep{brow01}. The 1$\sigma$ uncertainties were estimated by a bootstrap analysis based on simulated data sets constructed from the best fit residuals. Building each data set required: (i) to subtract the best fit model to the data, (ii) to sort again a fraction of $1/e$ of the residuals, (iii) to add again the best fit model to the new residuals. AMOEBA was used to fit all the parameters in each of the new data sets. The uncertainties were estimated as the standard deviation of each of the fitted parameters (see Table \ref{tab:par}).
\section{Follow-up observations}
\label{sec:fup}
\begin{figure}
\begin{center}
\epsfig{file=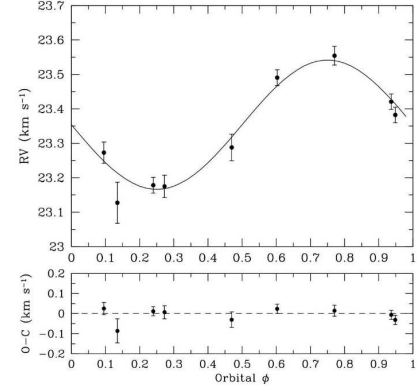,height=6.3cm,angle=0}
\caption{Radial-velocity variations of \exo1b versus phase from CoRoT's ephemeris. Top: the data fitted with a circular orbit of semi-amplitude K=$188 \ms$ and a drift of $1.02 \ms$ per day; bottom: the O-C residuals to the fit.}
\label{fig:figRV}
\end{center}
\end{figure}

\begin{figure}
\begin{center}
\epsfig{file=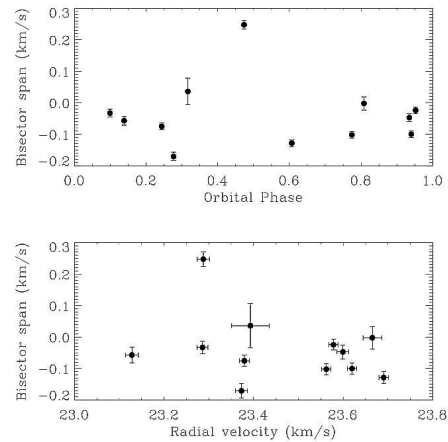,height=6.cm,angle=0}
\caption{Bisector spans: no variation with amplitude and phase correlates with the radial velocities  \citep[e.g.][]{que01}}
\label{fig:figRVbis}
\end{center}
\end{figure}

Observations of \exo1 were reported in the archive of the pre-launch photometric survey performed with the Berlin Exoplanet Search Telescope (BEST); the data however, were too noisy to clearly identify a transit egress. Follow-up photometry with the Wise observatory 1.0 m telescope confirmed that a transit occurs on the main target. Images taken at the CFHT showed only some faint background stars (up to V=22) that confirm the weak contamination inside CoRoT's photometric mask. The results of these investigations, indicative that none of the \exo1's near neighbour can be responsible for the transit signal, will be published elsewhere. High precision radial velocity observations of \exo1 were made at Observatoire de Haute Provence in March-April and October 2007 with SOPHIE, an echelle cross-dispersed fiber-fed spectrograph at the 193 cm telescope \citep{bou06}. The spectrograph routinely gets to a precision less than 2-3$\ms$ with the cross-correlation technique \citep{bara96} on bright targets. CoRoT-Exo-1 was observed with the High Efficiency mode (R=40,000) at an accuracy of about 20-30$\ms$ in typical exposure times of 45 min. Data reduction was made online and cross-correlation performed with a mask corresponding to a G2 star. The observations polluted by Moon illumination have been corrected for the correlation peak of the sky background, which is measured simultaneously through a neighbour fiber. The estimated error bars take into account such residual systematics that affect the Doppler measurements on faint targets, with signal-to-noise ratios ranging from 20 to 27.
\onltab{1}{
\begin{table}
\begin{center}
\caption{Radial velocity measurements of \exo1b obtained by SOPHIE associated to the system orbital phases (the large error bar of the second data point, due to acquisition problems, do not change significantly the orbital fitting).}
\label{tab:rad}
\begin{tabular}{lllll}
\hline
Julian & Orbit & RV & $\sigma_{RV}$  \\
date (d)& Phase & $ {\rm km s^{-1}}$  & $ {\rm km s^{-1}}$ \\
\hline
2454184.30498  & 0.469  &  23.2879  &  0.0385  \\
2454185.30867  & 0.134  &  23.1284  &  0.0600  \\
2454192.30397  &  0.770 &  23.5625  &  0.0271  \\
2454197.32090  &  0.095 &  23.2862  &  0.0309  \\
2454376.66501  &  0.948 &  23.5784  &  0.0226  \\
2454378.66336  &  0.273 &  23.3730  &  0.0324  \\
2454379.66472  &  0.936 &  23.6200  &  0.0224  \\
2454379.67016  &  0.602 &  23.6907  &  0.0229  \\
2454379.63151  &  0.240 &  23.3795  &  0.0227  \\
\hline
\end{tabular}
\end{center}
\end{table}
}
In total, 9 measurements were performed with SOPHIE and are reported in Table \ref{tab:rad} where are displayed: Julian date, planetary orbital phase, radial velocity and error. The measurements show a variation in phase with the ephemeris constrained, with a very high accuracy, by the CoRoT light curve. The orbital period and the epoch of the transits are fixed to the CoRoT values. As the period is very short, tidal circularization justifies a zero eccentricity assumption. The semi-amplitude of the radial velocity variation, K =188$\pm11\ms$, is compatible with a companion of planetary mass and was adjusted to the data (see Figure \ref{fig:figRV}). The best fit is obtained assuming a drift of 200$\ms$ between the two epochs of SOPHIE's observations or, equivalently, a linear drift of $1\ms $ per day. The final solution is displayed on Table \ref{tab:par}; the O-C residuals  have a standard deviation of 34\ms. The measurements do not correlate with the bisector of the cross-correlation function (see Figure \ref{fig:figRVbis}). In summary, the radial-velocity curve supports the planetary nature of the transiting body detected by CoRoT and discards other interpretations involving background stars, grazing eclipsing binary or a triple system. Additional observations are ongoing with HARPS to clarify the nature of the drift, possibly due to a second companion.
\section{Star and planet parameters}
\label{sec:params}
To derive the star's parameters, we took advantage of the spectroscopic observations performed to get the radial velocity measurements. The very first observations recorded with HARPS, whose resolving power is $R\sim100,000$, were used to perform a spectral analysis.  For each set of spectra, we co-added the individual exposures, order per order, once calibrated in absolute wavelength and rebinned  at a constant step of 0.02\AA. The resulting signal-to-noise ratio per pixel at 550 nm is $\sim80$ for the co-added HARPS spectra. The atmospheric parameters, \teff, \logg and \gmet~were derived by comparing the observed normalized spectrum to a library of synthetic spectra calculated with LTE MARCS models \citep{gus05}. The library sampled the range of temperature from 4000~K and 8000~K, surface gravity from \logg = 1.0 to 5.0 and metallicity from $-3.0$ up to $5.0$. The temperature was first estimated on the H$\alpha$ line wings, which is insensitive to any stellar parameters but the temperature. Special care has been devoted to the normalization of the corresponding order which could be quite tricky due to the extended wings of line. For this line, we used the new grids of ATLAS9 models \citep{heit02}. This value was controlled using the relative intensity of iron lines, from which a similar temperature was derived. Besides, this is also in good agreement with that estimated from its color index in the ExoDat database (B-V = 0.573) and the standard photometric calibrations. Metallicity is difficult to determine in this part of the spectrum as all indicators are also sensitive to gravity. Both were  estimated jointly by comparing the observed spectrum with the synthetic ones, until the best fit was achieved.  In our modeling of the spectra, we used the rotational velocity \vsini=5.2$\pm$1.0\kms derived from the widths of the HARPS cross-correlation function \citep{sant02}. Though more detailed studies are needed to achieve a precise determination, we find that \exo1 is similar to the Sun with a temperature of \teff = 5950K $\pm$150~K,  \logg =4.25$\pm0.30$ but a metal deficiency \gmet =$-0.30\pm0.25$. This metallicity is in agreement with the [Fe/H] value of $-0.4$ deduced from the calibration of the HARPS cross-correlation function and the star's color index \citep{sant02}. The stellar mass was derived using the evolutionary tracks \citep{seis00,more07} in the ($M_\star^{1/3}$/$R_{\star}$,\teff) plane. The method allows us to take advantage of the $M_\star^{1/3}/R_{\star}$ value measured on the CoRoT light curve with a very good precision, and to get rid of the large uncertainty on the gravity estimate. It will be detailed in a next paper devoted to the fundamental parameters of the first CoRoT planet host stars, based on forthcoming {\sl UVES} spectra. It is possible to assess that this star is quite an old main sequence star. The uncertainty on the mass, linked to the uncertainties on \teff~and [M/H], will be significantly reduced using complementary spectroscopic observations. We get $M_\star$=$0.95\Msun$ $\pm0.15$\Msun. From the planet side, only the uncertainty on the stellar mass limits our knowledge of the stellar and planet parameters \citep{brow01}.
\begin{table}
\begin{center}
\caption[]{The star and the planet parameters.}
\label{tab:par}
\begin{tabular}{lll}
{\bf Star parameters} &  \\
\hline
CorotID  &  0102890318   \\
RA $[J2000]$  & 6$^{h}$  48$^{m}$ 19.17$^{s}$  \\
Dec $[J2000]$ &  -3$^{o}$ 06' 07.78''  \\
Vmag  &  13.6   \\
$V_{0}$ $ {\rm [km/s]}$  &  23.354   $\pm0.008$ \\
K  $ {\rm [m/s]}$  & 188  $\pm11$ \\
v sin i  $ {\rm [km/s]}$ & 5.2 $\pm 1.0$  \\
\teff $ {\rm [K]}$ & 5950  $\pm150$ \\
\logg & 4.25 ${\bf \pm0.30}$ \\
$[M/H]$ & -0.3 $\pm0.25$ \\
$M_{\star}$ $[\Msun]$ & 0.95  $\pm$0.15\\
$R_{\star}$ $[\Rsun]$  & 1.11  $\pm$0.05 \\
 & \\  {\bf Planet parameters} &  \\
\hline
$k=R_{P}/R_{\star}$ &  0.1388 $\pm$0.0021\\
$T_{c}$ $ {\rm [d]}$  &  2454159.4532 $\pm$ 0.0001 \\
$M_{\star}^{1/3}/R_{\star}$  & 0.887 $\pm$0.014 \\
$u_{+}$  & 0.71 $\pm$0.16 \\
$u_{-}$   & 0.13 $\pm$0.30  \\
$P$ $ {\rm [d]}$   & 1.5089557 $\pm$ 0.0000064\\
a/$R_{\star}$ &  4.92 $\pm$0.08 \\
e  & 0 (fixed) \\
$i$ $ {\rm [deg]}$ & 85.1 $\pm0.5$ \\
$M_{P}$ $[\Mjup]$  & 1.03 $\pm$0.12 \\
$R_{P}$ $[\Rjup]$ & 1.49 $\pm$0.08 \\
$\rho_{P}$  $ {\rm [g~cm^{-3}}]$ & 0.38 $\pm$0.05 \\
$T_{P}$ $ {\rm [K]}$  & 1898 $\pm$50
\end{tabular}
\end{center}
\end{table}
\section{Conclusion}
\label{sec:conc}
\exo1b is a giant short-period planet orbiting at $\sim5$ stellar radii from a G0V star that seems to be metal-poor (further studies are necessary to confirm the metallicity). Its main characteristics are a large radius and a very low value of the mean density that may be consistent with a planet metal deficiency \citep{bur07}. The low mean density could be interpreted as planet inflation under strong irradiation and high loss rate \citep{lam03}. Similarly to  HD209458b, the planet is too large to be reproduced by standard evolution models. Its radius could be explained by advocating an additional heat source ($\sim 1\%$ of the incoming stellar flux) or an increase of the 2000-3000K opacities \citep[e.g.][]{guil06}. The star and the planet are in strong tidal interaction with a Doodson constant of the same order of magnitude as for OGLE-TR-56b \citep{paet04}. Due to its high photometric accuracy CoRoT will continue to help us improve our understanding of strongly irradiated planets. CoRoT-Exo-1b was detected early in the processing of the data. A lower level of noise with completely reduced data will open up the possibilities to search for outer planets, using transit timing variations, or to detect planet reflected light.

\begin{acknowledgements}
We warmly thank the OHP staff for his help observing with SOPHIE and the exoplanet consortium for the offered flexibility and the service observations (T. Forveille and C. Lovis). The austrian team thanks ASA for funding the CoRoT project. The team from the IAC acknowledges support by grants ESP2004-03855-C03-03 and ESP2007-65480-C02-02  of the Spanish Education and Science ministry. The German CoRoT team (TLS and Univ. Cologne) acknowledges the support of DLR grants 50OW0603, and 50QP0701.
\end{acknowledgements}
\bibliographystyle{aa}
\bibliography{references}

\begin{thebibliography}{}
\bibitem[Auvergne et al. ]{auv08} Auvergne, M. et al.  \aap, (in preparation)
\bibitem[Baglin et al. 2006]{bag06} Baglin, A. et al., 2006, 36th COSPAR ScientiÞc Assembly, 36, 3749
\bibitem[Baranne et al. 1996]{bara96} Baranne, A., Queloz, D., Mayor, M., et al. 1996, \aaps, 119, 373
\bibitem[Bouchy et al. 2006]{bou06} Bouchy, F. et al., 2006, in \textit{Tenth Anniversary of 51 Peg-b}, 319
\bibitem[Brown et al. 2001]{brow01} Brown, T., Charbonneau, D., Gilliland, R. et al. 2001, \apj,  552, 699
\bibitem[Burrows et al. 2007]{bur07} Burrows, A. et al., 2007, \apj, 661, 502
\bibitem[CoRoT 2006]{book06} CoRoT 2006
\textit{The CoRoT Mission: pre-launch status} {ESA-SP 1306}
\bibitem[Gim\'enez 2006] {gim06} Gim\'enez A. 2006,  \aap, 450, 1231
\bibitem[Guillot et al. 2006]{guil06} Guillot, T. et al.  2006, \aap,  453, L21
\bibitem[Gustafsson et al. 2005]{gus05} Gustafsson et al.  2005,
IAU Symp. 228, Cambridge Univ. Press, 259
\bibitem[Heiter et al. 2002]{heit02} Heiter et al., 2002, \aap, 392, 619
\bibitem[Lammer et al. 2003]{lam03} Lammer, H. et al., 2003, \apj, 598, L121
\bibitem[Morel \&  Lebreton 2007]{more07} Morel, P. \& Lebreton, Y.  2007, \apss, .tmp.,  460
\bibitem[P\"atzold et al. 2004]{paet04} P\"atzold, M., Carone, L.  \& Rauer, H.  2004, \aap,  427, 1075
\bibitem[Press et al. 1992]{pre92} Press, W. H. et al.  1992, Numerical recipes, Cambridge Univ. Press
\bibitem[Santos et al. 2002]{sant02} Santos, N. et al., 2002, \aap, 392, 215
\bibitem[Siess et al. 2000]{seis00} Seiss et al., 2000, 358, 593
\bibitem[Queloz et al 2001]{que01} Queloz, D. et al., 2001, \aap, 379, 279
\end{thebibliography}

\end{document}